\documentclass{article}
\usepackage{spconf,amsmath,graphicx}
\usepackage{amsfonts}
\usepackage{psfrag}
\usepackage{epsf} 
\usepackage{dsfont}
\usepackage{textcomp}
\usepackage{stmaryrd}
\usepackage{amssymb}
\usepackage{mathrsfs}
\usepackage{textcomp}
\usepackage{cite}
\usepackage{float}
\newcommand{\tr}[1]{\mathrm{#1}}
\usepackage[belowskip=-6pt,aboveskip=4pt,footnotesize]{caption}

\newcommand\blfootnote[1]{%
  \begingroup
  \renewcommand\thefootnote{}\footnote{#1}%
  \addtocounter{footnote}{-1}%
  \endgroup
}

\title{Effect of Synchronizing Coordinated Base Stations \\on Phase Noise Estimation}
\name{M. Reza Khanzadi$^{\textrm{\dag,*}}$, Rajet Krishnan$^{\textrm{*}}$, and Thomas Eriksson$^{\textrm{*}}$}
\address{}
\begin{document}
%
\maketitle
\blfootnote{\hspace{-0.5cm}\textdagger{Department of Microtechnology and Nanoscience, MEL,\\
\textasteriskcentered{Department of Signals and Systems, Communication Systems Group}, {Chalmers University of Technology, Gothenburg, Sweden.}
\\Email:\{khanzadi, rajet, thomase\}@chalmers.se}}
\vspace{-0.5cm}
\begin{abstract}
In this paper, we study the problem of oscillator phase noise (PN) estimation in coordinated multi-point (CoMP) transmission systems. Specifically, we investigate the effect of phase synchronization between coordinated base stations (BSs) on PN estimation at the user receiver (downlink channel). In this respect, the Bayesian Cram\'{e}r-Rao bound for PN estimation is derived which is a function of the level of phase synchronization between the coordinated BSs. Results show that quality of BS synchronization has a significant effect on the PN estimation.
\end{abstract}
\vspace{-0.3cm}
\section{Introduction}
\label{sec:intro}
Coordinated multi-point (CoMP) transmission is an approach to increase data transmission rate and improve quality of service in modern cellular communication networks \cite{Parkvall2008,Sawahashi2010,Irmer2011}. With CoMP, data is transmitted jointly from multiple coordinated base stations (BSs) at the same time, thereby improving the quality of the received signal at the user receiver \cite{Parkvall2008,Jingya2012}. 

One of the major challenges in CoMP joint transmission is carrier phase (and frequency) synchronization \cite{Jungnickel2008,Gesbert2010,Kotzsch2010,Irmer2011,Jingya2012}.  The synchronization problem in CoMP is two-fold. First, radio frequency local oscillators (LOs) at the coordinated BSs must be synchronized. Second, LO at the user receiver must be synchronized with those at the BSs. In general, oscillator phase noise (PN) evolves fast and it is not possible to fully synchronize the BSs only by exchanging the backhaul information \cite{Tolli2011,Krishnan2012_1}. To achieve an acceptable level of phase synchronization, very low-phase-noise LOs must be employed, which may be cost inefficient \cite{Jungnickel2008,Irmer2011}. An alternative approach is to track the overall PN at the user receiver. 

Carrier phase synchronization in single input-single output (SISO) and multiple input-multiple output (MIMO) systems has been extensively studied in the literature (e.g., \cite{Meyr1997,Dauwels2004,Schenk2005,Liu2006,Pedersen2008,Bhatti2009,Krishnan2012_1, Mehrpouyan2012,Khanzadi2011, Krishnan2011_1} and references therein). The effect of PN on the performance of MIMO networks when transmitters are not synchronized has been studied in \cite{Krishnan2012_1}. Authors in \cite{Mehrpouyan2012} have studied bounds on the performance of PN estimators in a similar MIMO setup. 

\begin{figure}[t]
\centering
\psfrag{LO}[cc][][0.6]{LO}%
\psfrag{Bs1}[cc][][0.6]{BS1}%
\psfrag{Bs2}[cc][][0.6]{BS2}%
\psfrag{UE}[cc][][0.7]{user}%
\psfrag{h1}[cc][][0.7]{$h_1$}%
\psfrag{h2}[cc][][0.7]{$h_2$}%
\psfrag{bdu}[cc][][0.5]{baseband}%
\psfrag{bd1}[ll][][0.6]{baseband}%
\psfrag{signal1}[ll][][0.6]{signal}%
\psfrag{signal2}[rr][][0.6]{signal}%
\psfrag{bd2}[rr][][0.6]{baseband}%
\psfrag{Los}[cc][][0.6]{LOs}%
\psfrag{synch}[cc][][0.6]{synchronization}%
\psfrag{link}[cc][][0.6]{link}%
\includegraphics [width=2.4in]{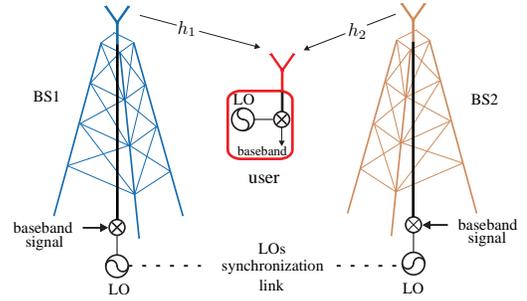}
\caption{Structure of the studied two BSs-one user CoMP system.}
\label{fig:1}
\vspace{-0.3cm}
\end{figure}

In this paper, we study how PN estimation in the downlink of CoMP joint transmission systems (i.e., at the user's receiver) is affected by synchronizing the BSs. To do so, we introduce a synchronization factor that models various levels of phase synchronization between the BSs. Then, we derive Bayesian Cram\'{e}r-Rao bound (BCRB) on the performance of data-aided (DA) and non-data-aided (NDA) PN estimators in a setup with two BSs and one user receiver (Fig.~\ref{fig:1}). We show that PN, when LOs are not phase synchronized, not only results in the phase rotation of received signal, but also leads to amplitude error that is analytically characterized in this work. Finally, we verify our results for various PN variances and synchronization factors by means of simulations.\footnote{{\bf Notations:} Italic letters $(x)$ are scalar variables, bold letters $(\mathbf{ x})$ are vectors, bold upper case letters $(\mathbf{X})$ are matrices, $([\mathbf{X}]_{a,b})$ denotes the $(a,b)^{th}$ entry of matrix $\mathbf{X}$, $\mathds{E}[\cdot]$ denotes the statistical expectation, $\Re(\cdot)$, $\Im(\cdot)$, and $\angle(\cdot)$ are real part, imaginary part, and angle of complex values, $(\cdot)^*$, and $(\cdot)^T$ denote conjugate and transpose, respectively and $\Delta_{\mathbf{x}}^{\mathbf{x}}(\cdot)$ denotes the second derivative with respect to vector $\mathbf{x}$.} 
\vspace{-0.3cm}
\section{SYSTEM MODEL}
\label{sec:system_model}
Consider the joint transmission of a sequence of complex-valued modulated symbols $\mathbf{s}=[s_1,\dots,s_N]$ in a two-BSs CoMP system. The digital base-band received signal at the user's receiver  $\mathbf{y}=[y_1,\dots,y_N]$  is modeled by
\vspace{-0.3cm}
\begin{align}
\label{system_model}
y_n=s_n\big(h_1e^{j(\overbrace{\phi^{t_1}_n+\phi^{r}_n}^{\triangleq\phi^{[1]}_n})}+h_2e^{j(\overbrace{\phi^{t_2}_n+\phi^{r}_n}^{\triangleq\phi^{[2]}_n})}\big)+w_n,
\end{align}
where $n=\{1,\dots,N\}$, $s_n$ is the $n^{th}$ complex modulated transmitted symbol from constellation $\mathcal{C}$, and $h_1$ and $h_2$ are the quasi-static channel gains from the BSs' antennas to the receiver's antenna. PN samples at the output of LOs at the first BS, second BS and receiver are denoted as $\phi^{t_1}_n$, $\phi^{t_2}_n$ and $\phi^{r}_n$, respectively. Further, $w_n$ denotes the zero-mean circularly symmetric complex-valued additive white Gaussian noise (AWGN) with variance $\sigma_w^2$ at the user's receiver. It should be noted that channel state information (CSI) is shared among the coordinated BSs \cite{Jingya2012}. Thus, in our study $h_1$ and $h_2$ are known and set to be unity. It is straightforward to generalize the results to the scenario where $h_1$ and $h_2$ can take any arbitrary value. Furthermore, in this paper we study the case where PN evolves much faster than the channel fading, which is in line with the assumptions in \cite{Darryl2007,Krishnan2012_1,Mehrpouyan2012,Jingya2012}.

PN is modeled as Wiener processes 
\begin{align}
\label{PN_model}
\phi^{i}_n=\phi^{i}_{n-1}+\zeta^{i}_{n-1},i\in \{t_1,t_2,r\},
\end{align}
where $\zeta^{i}_{n-1}$ is the PN innovation process with zero-mean Gaussian distribution and variance $\sigma^2_{\zeta}$ \cite{McNeill_thesis,Demir2000,Khanzadi2012_1}. 
The phase synchronizing the LOs at the BSs leads to correlated PNs which are modeled as two correlated Wiener processes.  That is, we set the PN innovations $\zeta^{t1}_{n-1}=\tilde{\zeta}^{t1}_{n-1} $ and $\zeta^{t2}_{n-1}=\rho\tilde{\zeta}^{t1}_{n-1}+\sqrt{1-\rho}\tilde{\zeta}^{t2}_{n-1}$, where $ \tilde{\zeta}^{t1}_{n-1} $ and $\tilde{\zeta}^{t2}_{n-1} $ are two independent zero-mean Gaussian processes with variance $\sigma^2_{\zeta}$, $\rho\in[0,1]$ denotes the correlation coefficient, and it can be shown that $\tr{Cov}(\zeta^{t1}_{n-1},\zeta^{t2}_{n-1})=\rho\sigma^2_{\zeta}$ \cite{platen2009}. Setting $\rho=1$ results in $\zeta^{t2}_{n-1}=\zeta^{t1}_{n-1}$, the fully phase synchronized LOs at the BSs, while $\rho=0$ leads to uncorrelated PN innovations that models unsynchronized LOs. In order to study the effect of PN on the received signal, we are interested in the performance of estimator of $\boldsymbol {\phi}=[\phi^{[1]}_{1},\dots,\phi^{[1]}_{N},\phi^{[2]}_{1},\dots,\phi^{[2]}_{N}]^T$, where $\phi^{[1]}_n$ and $\phi^{[2]}_n$ are defined in (\ref{system_model}). 

\section{Review of Bayesian Cram\'{e}r-Rao bound}
\label{ssec:Bound_BG}
Bayesian Cram\'{e}r-Rao bound (BCRB) gives a tight lower bound on the mean square error (MSE) of random parameter estimation \cite{VanTrees68} and it satisfies the following inequality:
\begin{align}
\label{BCRB_def}
&\mathbb{E}_{\mathbf{y},\boldsymbol{\phi}}\left[\left(\hat{\boldsymbol{\phi}}(\mathbf{y})-\boldsymbol{\phi}\right)\left(\hat{\boldsymbol{\phi}}(\mathbf{y})-\boldsymbol{\phi}\right)^T\right]\geq \mathbf{B}^{-1},\nonumber\\
&\mathbf{B}=\mathbb{E}_{\boldsymbol{\phi}}\left[\mathbf{F}(\boldsymbol{\phi})\right]+\mathbb{E}_{\boldsymbol{\phi}}\left[-\Delta_{\boldsymbol{\phi}}^{\boldsymbol{\phi}} \log f(\boldsymbol{\phi})\right],
\end{align}
where $\hat{\boldsymbol{\phi}}$ denotes an estimator of $\boldsymbol{\phi}$, $\mathbf{B}$ is the Bayesian information matrix (BIM), $f(\boldsymbol{\phi})$ is the prior distribution of $\boldsymbol{\phi}$, and $\mathbf{F}(\boldsymbol{\phi})$ is the so called Fisher information matrix (FIM) that for the DA case is defined as
\begin{align}
\label{FIM_DA}
&\mathbf{F}_\tr{DA}(\boldsymbol{\phi})=\mathbb{E}_{\mathbf{y}|\boldsymbol{\phi},\mathbf{s}}\left[-\Delta_{\boldsymbol{\phi}}^{\boldsymbol{\phi}} \log f(\mathbf{y}|\boldsymbol{\phi},\mathbf{s})\right].
\end{align}
For the NDA scenario, FIM is 
\begin{align}
\label{FIM}
&\mathbf{F}_\tr{NDA}(\boldsymbol{\phi})=\mathbb{E}_{\mathbf{y}|\boldsymbol{\phi}}\left[-\Delta_{\boldsymbol{\phi}}^{\boldsymbol{\phi}} \log f(\mathbf{y}|\boldsymbol{\phi})\right],
\end{align}
where a modified version of NDA FIM is
\begin{align}
\label{FIM_M}
&\mathbf{F}_\tr{MBCRB}(\boldsymbol{\phi})=\mathbb{E}_{\mathbf{s}}\left[\mathbb{E}_{\mathbf{y}|\boldsymbol{\phi},\mathbf{s}}\left[-\Delta_{\boldsymbol{\phi}}^{\boldsymbol{\phi}} \log f(\mathbf{y}|\boldsymbol{\phi},\mathbf{s})\right]\right],
\end{align}
that usually has a simpler analytical form compared to $\mathbf{F}_\tr{NDA}(\boldsymbol{\phi})$ and the corresponding bound is equivalently called modified Bayesian Cram\'{e}r-Rao bound (MBCRB) \cite{Andrea1994}.
\vspace{-0.4cm}
\section{Phase Noise Estimation in CoMP}
\label{sec:Bound_Calc}
\vspace{-0.3cm}
In this section, we derive analytical expressions of the DA and NDA BCRBs for PN estimation in CoMP. First, we find the terms involving in calculation of the BIM (\ref{BCRB_def}).
\vspace{-0.45cm}
\subsection{Calculation of $\mathbb{E}_{\boldsymbol{\phi}}\left[-\Delta_{\boldsymbol{\phi}}^{\boldsymbol{\phi}} \log f(\boldsymbol{\phi})\right]$}
\label{ssec:Prior}
According to (\ref{system_model}) and (\ref{PN_model}), we have
\begin{align}
\label{innovation_sum}
&\phi^{[1]}_n=\phi^{[t_1]}_1+\sum_{i=1}^{n-1}\zeta^{[t_1]}_i+\phi^{[r]}_1+\sum_{i=1}^{n-1}\zeta^{[r]}_i,\nonumber\\
&\phi^{[2]}_n=\phi^{[t_2]}_1+\sum_{i=1}^{n-1}\zeta^{[t_2]}_i+\phi^{[r]}_1+\sum_{i=1}^{n-1}\zeta^{[r]}_i,
\end{align}
where initial PN parameters $\phi^{[t_1]}_1$, $\phi^{[t_2]}_1$ and $\phi^{[r_1]}_1$ are modeled as zero-mean Gaussian random variables with a high variance\footnote{We consider a flat non-informative prior \cite{book_kay_est,Bay2008} for the initial PN values. To simplify the derivations, it is modeled by a Gaussian distribution with a high variance that behaves similar to a flat prior over a certain interval.}, denoted as $\sigma^2_1$. Based on (\ref{innovation_sum}), we can show that $\boldsymbol {\phi}$ has a multivariate Gaussian distribution $f(\boldsymbol {\phi})=\mathcal{N}(\boldsymbol {\phi};\mathbf{0},\mathbf{C})$ where the covariance matrix $\mathbf{C}$ follows the form
\vspace{-0.1cm}
\begin{align}
\label{C_def}
\mathbf{C}=\left[ \begin{array}{c|c}
\mathbf{C}_1 & \mathbf{C}_2 \\ \hline
\mathbf{C}_3 & \mathbf{C}_4 \end{array} \right],
\end{align}
\vskip -0.3cm \hskip -0.5cm and 
\begin{align}
&[\mathbf{C}_1]_{l,k}=[\mathbf{C}_4]_{l,k}=2(\sigma^2_0+\sigma^2_{\zeta}\min(l-1,k-1)),\nonumber\\
&[\mathbf{C}_2]_{l,k}=[\mathbf{C}_3]_{l,k}=(1+\rho)(\sigma^2_0+\sigma^2_{\zeta}\min(l-1,k-1)),\nonumber\\
&l,k\in\{1,\dots,N\}.
\end{align}
Based on definition of $f(\boldsymbol {\phi})$, it is straightforward to show that
\begin{align}
\mathbb{E}_{\boldsymbol{\phi}}\left[-\Delta_{\boldsymbol{\phi}}^{\boldsymbol{\phi}} \log f(\boldsymbol{\phi})\right]=\mathbf{C}^{-1}.
\end{align}
\vspace{-.8cm}
\subsection{Calculation of $\mathbb{E}_{\boldsymbol{\phi}}\left[\mathbf{F}(\boldsymbol{\phi})\right]$}
\label{ssec:Likelihood}
In order to obtain FIM, we need to compute the likelihood function. When $\mathbf{s}$ is known, the likelihood function reads
\begin{align}
\label{DA_Likelihood}
\hspace{-0.3cm}f(\mathbf{y}|\boldsymbol {\phi},\mathbf{s})&=\prod_{n=1}^Nf(y_n|\phi^{[1]}_{n},\phi^{[2]}_{n},s_n)\nonumber\\
&=\left(\frac{1}{\sigma_w^2\pi}\right)^N\prod_{n=1}^Ne^{-\frac{|y_n|^2+2|s_n|^2(1+\cos(\phi^{[1]}_{n}-\phi^{[2]}_{n}))}{\sigma_w^2}}\nonumber\\
&\hspace{2.1cm}\times e^{\frac{2}{\sigma^2_w}\Re\{y_n s_n^*(e^{-j\phi^{[1]}_{n}}+e^{-j\phi^{[2]}_{n}})\}}.
\end{align}
To find the likelihood when $\mathbf{s}$ is not known, one needs to take the expectation of the likelihood at time $n$ with respect to all possible transmitted symbols  
\begin{align}
\label{NDA_Likelihood}
\hspace{-0.3cm}f(\mathbf{y}|\boldsymbol {\phi})&=\prod_{n=1}^N\sum_{s_n\in \mathcal{C}} \frac{1}{M}f(y_n|\phi^{[1]}_{n},\phi^{[2]}_{n},s_n)\nonumber\\
&=\left(\frac{1}{\sigma_w^2\pi}\right)^N\prod_{n=1}^N\sum_{s_n\in \mathcal{C}} \frac{1}{M} e^{-\frac{|y_n|^2+2|s_n|^2(1+\cos(\phi^{[1]}_{n}-\phi^{[2]}_{n}))}{\sigma_w^2}}\nonumber\\
&\hspace{2.4cm}\times e^{\frac{2}{\sigma^2_w}\Re\{y_n s_n^*(e^{-j\phi^{[1]}_{n}}+e^{-j\phi^{[2]}_{n}})\}},
\end{align}
where $M$ is the constellation order. Using (\ref{DA_Likelihood}), one can show that for the DA BCRB and MBCRB cases
\vspace{-0.1cm}
\begin{align}
\mathbb{E}_{\boldsymbol{\phi}}\left[\mathbf{F}(\boldsymbol{\phi})\right]=\left[ \begin{array}{c|c}
\mathbf{\Gamma} & \mathbf{0} \\ \hline
\mathbf{0} & \mathbf{\Gamma} \end{array} \right],
\end{align}
where $\mathbf{\Gamma}$ is a diagonal matrix and
\begin{align}
\label{gammas_mtrix}
&\hspace{-0.2cm}[\mathbf{\Gamma}]_{n,n}=\mathbb{E}\left[-\frac{\partial^2 \log f(y_n|\phi^{[1]}_{n},\phi^{[2]}_{n},s_n)}{\partial\phi_n^{[1]}\partial\phi_n^{[2]}}\right].
\end{align}
The diagonal elements of $\mathbf{\Gamma}$ have the following analytical forms: 
\begin{align}
\label{gammas_DA}
&\tr{DA:}~[\mathbf{\Gamma}]_{n,n}=\frac{2|s_n|^2}{\sigma_w^2}&&\tr{MBCRB:}~[\mathbf{\Gamma}]_{n,n}=\frac{2E_s}{\sigma_w^2},
\end{align}
where $E_s$ is the average symbol energy of the constellation. For the standard NDA BCRB
\vspace{-0.1cm}
\begin{align}
\mathbb{E}_{\boldsymbol{\phi}}\left[\mathbf{F}(\boldsymbol{\phi})\right]=\left[ \begin{array}{c|c}
\gamma_{11}\mathbf{I} & \gamma_{12}\mathbf{I} \\ \hline
\gamma_{21}\mathbf{I} & \gamma_{22}\mathbf{I} \end{array} \right],
\end{align}
where 
\begin{align}
\label{gammas_NDA}
&\gamma_{ij}=\mathbb{E}\left[-\frac{\partial^2 \log f(y_n|\phi^{[1]}_{n},\phi^{[2]}_{n})}{\partial\phi_n^{[i]}\partial\phi_n^{[j]}}\right]&&i,j\in\{1,2\},
\end{align}
and $\mathbf{I}$ denotes the identity matrix. Derivatives in (\ref{gammas_NDA}) can be expressed as
\begin{align}
\label{NDA_Likelihood_Derivative_general}
&\frac{\partial^2 \log f(y_n|\phi^{[1]}_{n},\phi^{[2]}_{n})}{\partial\phi_n^{[j]}\partial\phi_n^{[i]}}\nonumber\\
&\hspace{0.3cm}=\frac{\displaystyle\sum_{s_n\in \mathcal{C}}\frac{\partial^2 f_n}{\partial\phi_n^{[j]}\partial\phi_n^{[i]}}}{\displaystyle\sum_{s_n\in \mathcal{C}}f_n}-\frac{\left(\displaystyle\sum_{s_n\in \mathcal{C}}\frac{\partial f_n}{\partial\phi_n^{[j]}}\right)\left(\displaystyle\sum_{s_n\in \mathcal{C}}\frac{\partial f_n}{\partial\phi_n^{[i]}}\right)}{\left(\displaystyle\sum_{s_n\in \mathcal{C}}f_n\right)^2},
\end{align}
where $f_n\triangleq f(y_n|\phi^{[1]}_{n},\phi^{[2]}_{n},s_n)$. Although the first and second derivatives of $f_n$ in (\ref{NDA_Likelihood_Derivative_general}) have analytical forms, (\ref{gammas_NDA}) does not have a general closed-from solution and $\gamma_{ij}$ must be computed by means of numerical methods. 

Now that terms contributing to BIM are computed for different scenarios, corresponding bounds can be obtained by inverting the BIMs. In the next section, we use the computed bounds to study the effect of residual PN estimation errors on the amplitude of the received signals.

\section{Residual Amplitude Noise}
\label{sec:AMP}
Study of the PN-affected SISO systems shows that PN only affects phase of the transmitted symbols \cite{Khanzadi2011,Krishnan2013_1}. On the other hand, in a CoMP system also the amplitude is affected. Consider a scenario where we use a PN estimator that reaches the derived Cram\'{e}r-Rao bounds. We define the estimated PNs at time $n$ as $ \hat{\phi}^{[1]}_{n}$ and $\hat{\phi}^{[2]}_{n}$, while the PN estimation errors are denoted as $\epsilon^{[1]}_n\triangleq(\phi^{[1]}_{n}-\hat{\phi}^{[1]}_{n})$ and $\epsilon^{[2]}_n\triangleq(\phi^{[2]}_{n}-\hat{\phi}^{[2]}_{n})$ and modeled as Gaussian random variables. The amplitude of the $n^{th}$ received signal, distorted by residual PN errors, can be written as
\begin{align}
\label{AMP_PN1}
|y_n|&=|s_n\left(e^{j\epsilon^{[1]}_n}+e^{j\epsilon^{[2]}_n}\right)+w_n|\nonumber\\
&=\Big||s_n||e^{j\epsilon^{[1]}_n}+e^{j\epsilon^{[2]}_n}|e^{j(\angle s_n+\angle(e^{j\epsilon^{[1]}_n}+e^{j\epsilon^{[2]}_n}))}+w_n\Big|\nonumber\\
&=\sqrt{(2|s_n|\cos(\tilde{\epsilon}_n)+\Re \{w'_n\})^2+\Im^2 \{w'_n\}}\nonumber\\
&\overset{\tr{(a)}}{\approx} 2|s_n|\cos(\tilde{\epsilon}_n)+\Re \{w'_n\}\nonumber\\
&\overset{\tr{(b)}}{\approx} 2|s_n|(1-\frac{\tilde{\epsilon}_n^2}{2})+\Re \{w'_n\}\nonumber\\
&= 2|s_n|-\sigma^2_{\tilde{\epsilon}_n} q_n |s_n|+\Re \{w'_n\},
\end{align}
where $\Re\{w'_n\}= \Re\{w_n e^{j(\angle s_n+\angle(e^{j\epsilon^{[1]}_n}+e^{j\epsilon^{[2]}_n}))}\}$ is a zero-mean real Gaussian random variable with variance $\sigma^2_n/2$, $\tilde{\epsilon}_n\triangleq (\epsilon_n^{[1]}-\epsilon_n^{[2]})/2$ has a Gaussian distribution $f(\tilde{\epsilon}_n)=\mathcal{N}(\tilde{\epsilon}_n;0,\sigma^2_{\tilde{\epsilon}_n})$, where $\sigma^2_{\tilde{\epsilon}_n}=(\sigma^2_{\epsilon^{[1]}_n}+\sigma^2_{\epsilon^{[2]}_n}-2\tr{Cov}(\epsilon^{[1]}_n,\epsilon^{[2]}_n))/4$  and its value can be computed from the inverse of BIM, and $q_n$ is a random variable with chi-squared distribution: $q_n\sim\chi^2_1$. Approximations (a) and (b) in (\ref{AMP_PN1}) are based on high signal to noise ratio (SNR) and small PN error assumptions, respectively which are validated in the simulation section. In (\ref{AMP_PN1}), $\sigma^2_{\tilde{\epsilon}_n} q_n |s_n|$ is the amplitude noise due to the PN estimation errors that has a chi-squared distribution and is a function of transmitted symbol's amplitude and $\sigma^2_{\tilde{\epsilon}_n}$. 
\section{Simulation Results}
\label{sec:Simulation Results}
\begin{figure}[t!]
\centering
\includegraphics [width=3.2in]{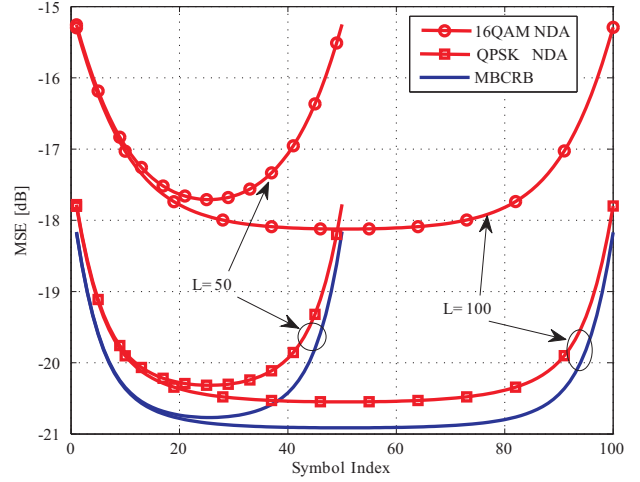}
\caption{BCRB vs. symbol index in a block, $\rho=0.5,~\sigma^2_\zeta=10^{-3},~\mathrm{SNR}=5~\mathrm{dB}$.}
\label{fig:2}
\vspace{-0.1cm}
\end{figure}

\begin{figure}[t!]
\centering
\includegraphics [width=3.2in]{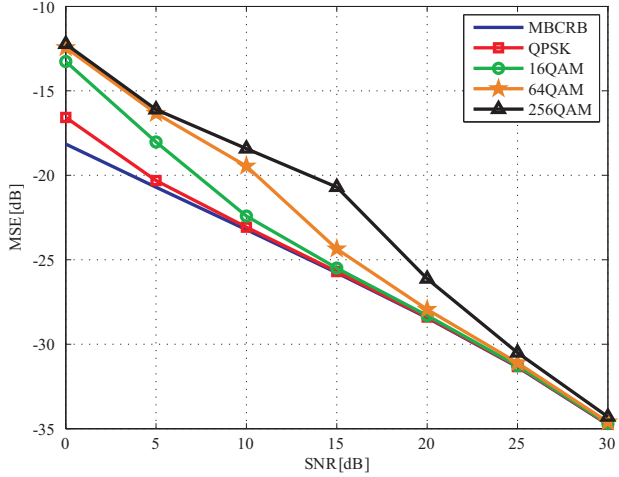}
\caption{BCRB vs. SNR, $L=100,~\rho=0.1,~\sigma^2_\zeta=10^{-3}$~rad$^2$.}
\label{fig:3}
\vspace{-0.1cm}
\end{figure}

\begin{figure}[t!]
\centering
\psfrag{rho}[cc][][1]{$\rho$}%
\includegraphics [width=3.2in]{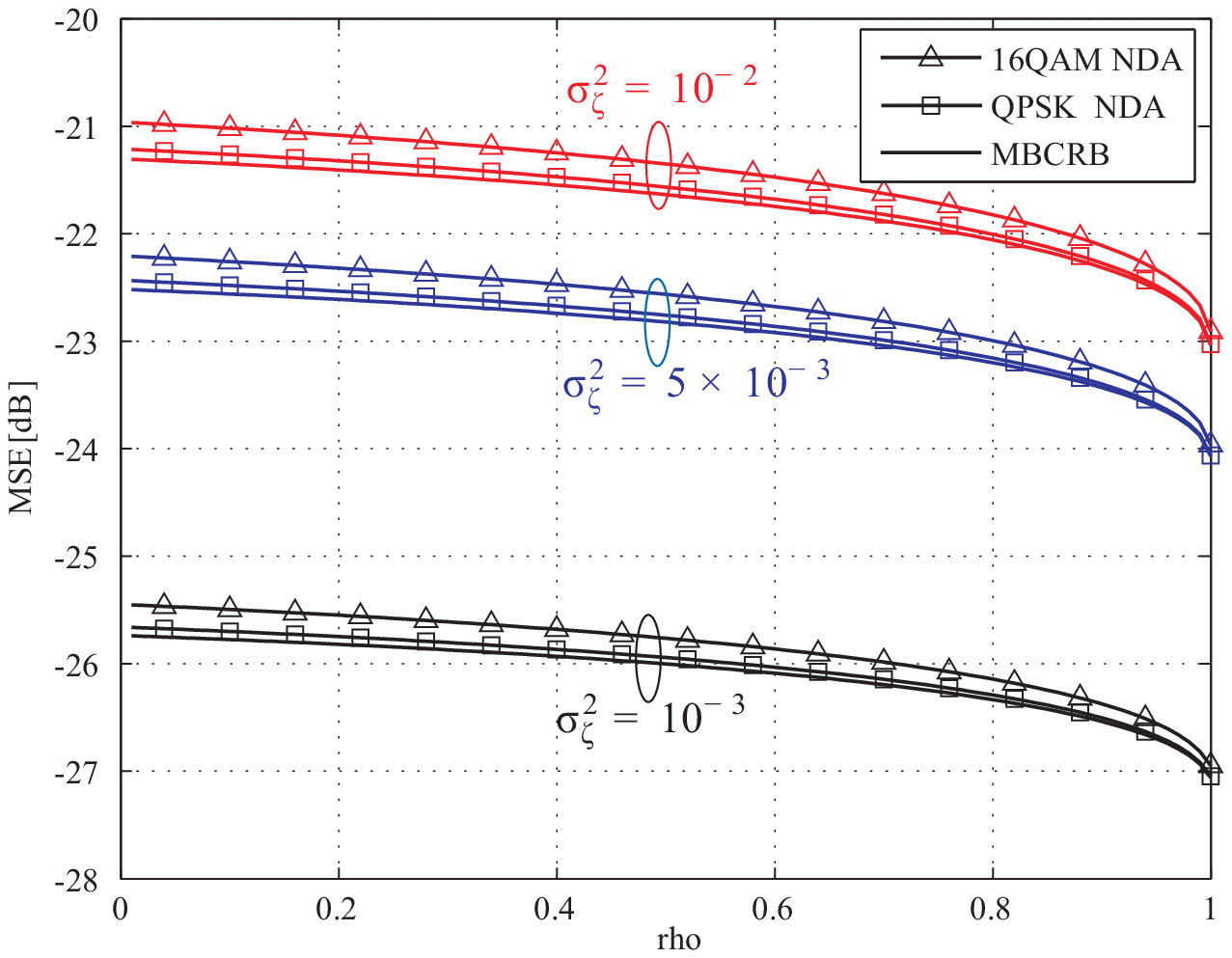}
\caption{BCRB vs. $\rho$, $\mathrm{SNR}=15~\mathrm{dB}$.}
\label{fig:4}
\vspace{-0.1cm}
\end{figure}

\begin{figure}[t!]
\centering
\psfrag{rho}[cc][][1]{$\rho$}%
\psfrag{s2eps}[cc][][1]{$\sigma^2_{\tilde{\epsilon}_n}$}%
\includegraphics [width=3.2in]{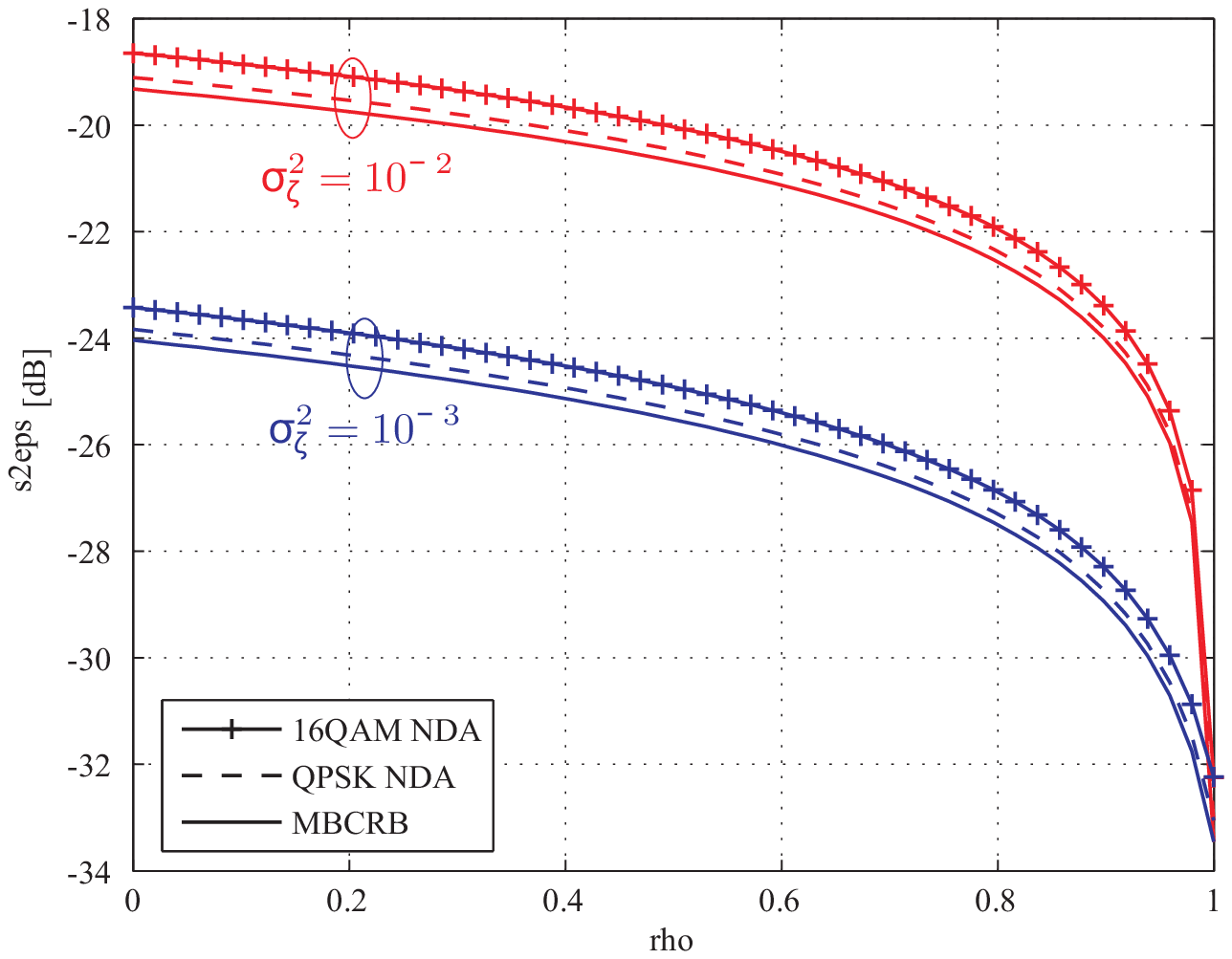}
\caption{$\sigma^2_{\tilde{\epsilon}_n}$ vs. $\rho$, $\mathrm{SNR}=15~\mathrm{dB}$.}
\label{fig:5}
\vspace{-0.1cm}
\end{figure}

In the following section, we present simulation results of the computed bounds for various signal constellations and other system parameters. According to (\ref{gammas_DA}), the DA BCRB is a function of the instantaneous amplitudes of the transmitted sequence of symbols.  Therefore, we only present the results for the NDA cases that depend on the average energy of the constellation. Expression (\ref{system_model}) shows that our system model for known channel gains is symmetric in $\phi^{[1]}_{n}$ and $\phi^{[2]}_{n}$ which results in identical estimation performance for these parameters. 

Fig.~\ref{fig:2} compares the behavior of standard NDA BCRB and MBCRB for estimation of $[\phi^{[1]}_{1},\dots,\phi^{[1]}_{N}]^T$ for various block lengths and constellations. This figure shows, although MBCRB has a simpler analytical form, it is a looser bound compare to the exact NDA BCRB. Moreover, NDA 16QAM has much higher MSE compared to NDA QPSK. This is due to the fact that by increasing the constellation order in the NDA case, the probability of making mistakes between the symbols increases that severely impacts the performance of PN estimation. In addition, Fig.~\ref{fig:2} shows that increasing the block length improves the PN estimation which is because of employing more observation symbols.

Fig.~\ref{fig:3} illustrates the MSE of estimating $\phi^{[1]}_{n=50}$ for different constellations as a function of SNR, where the block length is $N=100$. In general, by increasing the SNR, observations become more reliable and MSE of the NDA PN estimation gets smaller. This figure also shows that MBCRB is a tight bound for NDA PN estimation when SNR is high. Similar results has been reported for the SISO systems \cite{Moeneclaey98,Jianxiao2011}. 

Fig.~\ref{fig:4} depicts the effect of changing $\sigma^2_\zeta$ and $\rho$ on the PN estimation performance. First, it can be seen that by decreasing $\sigma^2_\zeta$ the MSE has been also reduced that follows our expectations; when PN evolves slower, it can be estimated more accurately. Second, this figure shows that by better synchronization of BSs (increasing $\rho$), the PN estimation performance is improved. This improvement is more prominent when LOs have higher PN variance. For example, when $\sigma^2_\zeta=10^{-3}$ rad$^2$ and BSs are fully synchronized ($\rho=1$), there is almost $1.5$~dB improvement in the MSE comparing to the unsynchronized case ($\rho=0$). However, this difference is $2$~dB when $\sigma^2_\zeta=10^{-2}$ rad$^2$. 

Fig.~\ref{fig:5} shows the effect of changing $\rho$ on the amplitude noise. By increasing $\rho$ to $1$, $\sigma^2_{\tilde{\epsilon}_n}$ tends to very small values that results in negligible amplitude noise. This result implies that increasing $\rho$ not only improves the PN estimation performance (see Fig. ~\ref{fig:4}), but also it increases the correlation between $\epsilon^{[1]}_n$ and $\epsilon^{[2]}_n$ that results in smaller $\sigma^2_\zeta$. 
\section{Conclusions And Future Work}
\label{sec:Conclusion}
In this paper, we have studied the effect of synchronization of coordinated BSs on the PN estimation at the user receiver. To this end, we have derived the DA and NDA BCRBs for the PN estimation in a CoMP system, valid for any arbitrary modulation format. Results show synchronization of BSs has a significant effect on the PN estimation performance and also the amplitude distortion of the received signal.  

To extend this work, it is possible to generalize the results to the case where arbitrary number of BSs cooperate. In such a scenario, if the BSs are not fully synchronized, the number of estimated parameters of interest increases that may result in a higher overall estimation error. Consequently, BSs synchronization would become more important in such scenarios.


\clearpage 
\small
\bibliographystyle{IEEEbib}
\bibliography{references}

\end{document}